# Evaluation of Spatial Resolution and Noise Sensitivity of sLORETA Method for EEG Source Localization Using Low-Density Headsets

S. Saha[1], Ya.I. Nesterets[2], M. Tahtali[3] and T.E. Gureyev[4]

**Abstract**  Electroencephalography (EEG) has enjoyed considerable attention over the past century and has been applied for diagnosis of epilepsy, stroke, traumatic brain injury and other disorders where 3D localization of electrical activity in the brain is potentially of great diagnostic value. In this study we evaluate the precision and accuracy of spatial localization of electrical activity in the brain delivered by a popular reconstruction technique sLORETA applied to EEG data collected by two commonly used low-density headsets with 14 and 19 measurement channels, respectively. Numerical experiments were performed for a realistic head model obtained by segmentation of MRI images. The EEG source localization study was conducted with a simulated single active dipole, as well as with two spatially separated simultaneously active dipoles, as a function of dipole positions across the neocortex, with several different noise levels in the EEG signals registered on the scalp. The results indicate that while the reconstruction accuracy and precision of the sLORETA method are consistently high in the case of a single active dipole, even with the low-resolution EEG configurations considered in the present study, successful localization is much more problematic in the case of two simultaneously active dipoles. The quantitative analysis of the width of the reconstructed distributions of the electrical activity allows us to specify the lower bound for the spatial resolution of the sLORETA-based 3D source localization in the considered cases.

## 1 Introduction

As a non-invasive modality for monitoring brain activity, Electroencephalography (EEG) has enjoyed considerable attention over the past century [1, 2]. In the most wide-spread of its forms, EEG measures the voltage potentials (in the order of micro-volts) at various locations on the scalp and employs signal processing techniques to infer the electrical activity inside the brain. Brain activity information accessible in EEG is largely complementary to that in MRI: EEG has excellent time resolution (down to a few milliseconds), but poor space resolution (few centimeters). In addition to the spectral content analysis of EEG recordings at multiple electrode locations on the scalp, spatial localization of the sources of electrical activity inside the brain has also been the subject of active research [3, 4]. EEG has been recently applied for diagnosis of epilepsy [3], stroke [5, 6], traumatic brain injury [7] and other disorders where 3D localization of abnormal electrical activity in the brain is potentially of great diagnostic value. Therefore, it is important and timely to quantify the precision and accuracy of 3D spatial localization provided by popular techniques such as sLORETA [8].

---

[1] S. Saha is a PhD student at the Department of Electrical Engineering in the University of New South Wales, Canberra, Australia and Intern in Biomedical Imaging at CSIRO Materials Science and Engineering, Melbourne, Australia
E-mail: S.Saha@student.adfa.edu.au, Sajib.Saha@csiro.au

[2] Ya.I. Nesterets is with CSIRO Materials Science and Engineering, Melbourne, Australia
E-mail: Yakov.Nesterets@csiro.au

[3] M. Tahtali is with the School of Engineering and Information Technology, University of New South Wales, Canberra, Australia
E-mail: M.Tahtali@adfa.edu.au

[4] T.E. Gureyev is a Senior Principal Research Scientist in CSIRO, Australia and an Adjunct Professor at the University of New England, Australia
E-mail: Tim.Gureyev@csiro.au

It is well established that neural electrical activity of the brain can be approximated in EEG by sets of virtual current dipoles [1, 2]. Source localization works by first finding the scalp potentials produced by virtual dipoles at arbitrary locations in the brain (i.e. solving the forward problem), then, in conjunction with the actual EEG data measured by the electrodes, it is used to work back and estimate the sources that best fit the measurements (i.e. solving the corresponding inverse problem). In the cases where the number of measurement points (electrodes) is lower than the number of unknowns (i.e. potential positions inside the head of the electrical dipoles with unknown current strength and orientation) this inverse problem is severely ill-posed [9, 10] in the sense that there is an infinite number of source configurations that can produce the same distribution of the electric potential on the surface of the head. Hence additional constraints need to be introduced in order to produce an appropriate unique solution. Note, however, that even with infinitely many data measurement points on the scalp, the spatial resolution of the EEG inversion will be limited due to the spreading of the electromagnetic signal on propagation through the head [2], which implies a non-uniqueness and/or severe instability of solutions to the source localization problem at high spatial resolutions.

Various methods have been proposed for choosing suitable constraints for the inverse EEG problem, the most well-known being the 'minimum norm' constraint [11, 12]. Techniques relying on the 'minimum norm' constraint are based on a search for the solution with minimum power, and correspond to Tikhonov regularization [9]. In other words, when the system is underdetermined, the solution is obtained by minimizing its L2-norm [13]. Several variants of this approach that consider different regularization parameters and weighting factors have already been proposed in the literature [9, 14]. The present paper is focused on the sLORETA method [8], which also relies on the electrical current density estimate given by the minimum norm solution and then "standardizes" it by using its expected variance [8]. The expected variance here is hypothesized to be due to the actual source variance ("biological noise") and the variance due to noisy measurements ("electronic noise") [14]. While in many cases sLORETA tends to produce very broadly distributed or "smeared" sources in the reconstruction region, it remains very popular among EEG researchers because of its reconstruction speed and its remarkable capacity to ensure zero localization error in the case of a single source and noiseless environment [17]. Even though the solution produced by sLORETA is typically blurry, when the sources are large in number and extended, sLORETA was found to be superior compared to some high resolution algorithms [15]. Besides, some high-resolution algorithms depend on low-resolution algorithms, such as sLORETA, for robust initialization. For instance, Shrinking LORETA-FOCUSS [16] and Standardized Shrinking LORETA-FOCUSS [15] are two particular examples where sLORETA has been used to ensure the robust initialization.

The present study aims at quantifying the localization error of sLORETA in the presence of noise, including the estimation of the spatial resolution of the reconstructed signal in the case of single dipole activation, and the localization error in the case of multiple dipole activation. In a paper by Wagner *et al.* [17], a similar study was conducted using a 'three spherical shells' model with 16,375 dipoles and a 81-electrode setup. Ding *et al.* in [18] compared sLORETA with several state-of-the-art methods in regard to localization error and orientation sensitivity. An experimental study was performed on a three-shell boundary element (BE) [19] model with 500 dipoles and a 32-channel electrode setup. In the present work we consider a realistic head model obtained by the segmentation of MRI images, with 6203 dipole locations uniformly distributed over the whole segmented volume of the neocortex, along with two different electrode setups (Fig.1) that correspond to two popular EEG headsets [20, 21]. Because the problem of reconstructing the 3D electrical activity of the brain from the signals collected on the surface of the head is highly underdetermined and ill-posed (as explained above), a small change in the considered model can potentially cause large changes in the solution. Realistic head models with high-density electrode setups have already been used in literature to investigate the reconstruction accuracy of sLORETA and related methods [15]. The present study is primarily aimed at a quantitative evaluation of the reconstruction accuracy that can be delivered by the sLORETA method applied to the EEG data collected by low-density EEG headsets, in the case of a single and multiple simultaneous sources of electrical activity in the brain and different noise levels in the data, including the estimation of the difference in the accuracy of the reconstruction results as a function of the number of electrodes in a low-density EEG setup.

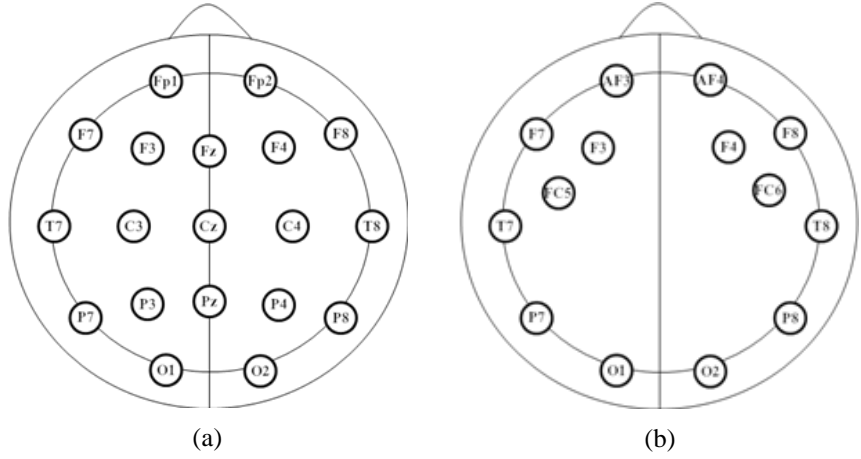

**Fig. 1** Schematic representation of the electrodes (a) Setup I (corresponds to a 19-channel Compumedics system) and (b) Setup II (corresponds to 14-channel Emotiv EPOC system).

## 2 Mathematical Formulation

The equation often used for defining the forward and inverse imaging problems in EEG has the following form:
$$\boldsymbol{\Phi} = \boldsymbol{KJ} + c\mathbf{1}. \tag{1}$$
Here $\boldsymbol{\Phi} \in \Re^{3N_E \times 1}$ is a vector of the scalp electric potentials measured by the $N_E$ electrodes with respect to a certain reference potential, $\boldsymbol{J} \in \Re^{3N_V \times 1}$ is the primary or impressed current density vector, where $N_V$ is the number of considered dipole locations in the brain, with each dipole current having three independent components corresponding to the usual Cartesian coordinates in 3D space, $\boldsymbol{K} \in \Re^{N_E \times N_{3V}}$ is the so-called lead field matrix, $c$ is a constant[5] and $\mathbf{1} \in \Re^{N_E \times 1}$ is a vector of ones. When the average reference transforms [8] of $\boldsymbol{\Phi}$ and $\boldsymbol{K}$ are used, equation (1) can be simplified:
$$\boldsymbol{\Phi} = \boldsymbol{KJ}. \tag{2}$$
Here the centering matrix, $\boldsymbol{H} = \boldsymbol{I} - \mathbf{11}^T / \mathbf{1}^T \mathbf{1}$ is used as the average reference operator and $\boldsymbol{I} \in \Re^{N_E \times N_E}$ is the identity matrix.

Equations (1) and (2) are severely under-determined with respect to the vector $\boldsymbol{J}$, as typically one has $N_E \ll N_V$. Hämäläinen *et al.* [22] were the first to propose a meaningful solution for this ill-posed problem. The minimum norm solution of eq.(2) in [22] was based on minimizing the cost function
$$\boldsymbol{F} = \|\boldsymbol{\Phi} - \boldsymbol{KJ}\|^2 + \alpha \|\boldsymbol{J}\|^2. \tag{3}$$
The vector $\hat{\boldsymbol{J}}$, which minimizes equation (3), was computed as $\hat{\boldsymbol{J}} = \boldsymbol{L\Phi}$ where $\boldsymbol{L} = \boldsymbol{K}^T [\boldsymbol{KK}^T + \alpha \boldsymbol{H}]^+$, where "+" denotes the Moore-Penrose pseudo inverse, and $\alpha$ is the regularization parameter. While the minimum norm solution is quite easy to compute, it is notorious for misplacing the deep sources

---
[5] Constant $c$ embodies the fact that an electric potential is determined up to an arbitrary constant. It allows the use of any reference for the lead field and measurements [8].

[8]. The method proposed by Dale *et al.* [23] enhances the minimum norm solution through the so-called "standardization" of the current density estimates. Nevertheless, Dale *et al.*'s method still produces a systematic non-zero localization error. The method subsequently proposed by Pascual-Marqui and named sLORETA [8] produces zero localization error (at least in the case of a single source and a noiseless environment). sLORETA relies on the following estimate of the standardized current density power [8]:

$$\hat{J}_l^T \{[S_{\hat{j}}]_{ll}\} \hat{J}_l. \tag{4}$$

Here $S_{\hat{j}} = L(KK^T + \alpha H)L^T = K^T(KK^T + \alpha H)^+ K.$

## 3 Robustness of sLORETA with Respect to Noise

Experiments were conducted on a realistic head model that was obtained by segmentation of MRI images of the head and included four different major components, namely scalp, skull, Cerebrospinal fluid (CSF) and brain with the following relative conductivity values [2]: $\sigma_{scalp}=1$, $\sigma_{skull}=0.05$, $\sigma_{CSF}=5$, $\sigma_{brain}=1$. The head model was defined on a discrete grid with a grid step of 1.25mm. The source space was constructed by dividing the head model into cubes with a side of 5mm and considering possible current dipoles only in those cubes that consisted of at least 60% gray matter. This segmentation procedure resulted in 6203 dipole positions. In general, dipoles can be of different strength and randomly oriented, however in the present experiment, we considered for simplicity that the active dipoles had unit length and were vertically oriented. Electrode locations corresponding to a 19-channel Compumedics (Setup I) and 14-channel Emotiv EPOC (Setup II) headsets were used for the study.

One of the aims of this work was to study the localization accuracy of sLORETA in regard to noise. When analysing real EEG data, the potentials measured on the scalp can be disturbed by several factors such as measurement noise, mislocalization of the electrodes, etc. [24, 25], which all can potentially induce errors in solutions of the source localization problem. Here we will investigate the effect of simulated pseudo-random measurement noise superimposed on the measured scalp potentials.

First, a single active dipole was considered and the corresponding potentials at the electrodes were calculated using eq.(2) with a lead-field matrix calculated by solving the direct EEG problem, i.e. by solving the corresponding second-order elliptical partial differential equation with piecewise-constant coefficients (corresponding to the constant conductivity values within each of the four segmented components of the head as described above). We then added variable amounts of noise to each of the calculated electrode potentials. Simulated, as well as realistic noise was considered in the experiment. The simulated random noise was modeled as statistically independent Gaussian noise with zero mean and standard deviation $\sigma$. For a given Signal-to-Noise Ratio (SNR), the standard deviation was computed as $\sigma = \max\{(P_{Signal}^e)\}/\text{SNR}$ where $P_{Signal}^e$ (in $\mu V$) was the signal measured by the electrode $e$ ($\forall_e \in 1:N_E$). The realistic noise was generated using real EEG data collected with an Emotiv EPOC headset [20]. In the latter case, 1000 consecutive samples of the EEG data samples from an interval showing no significant spikes were selected. For each electrode channel, the mean value of the 1000 data points was subtracted from the signal, and then the signal was re-scaled dividing it by its standard deviation. The new standard deviation was then imposed, depending on the desired SNR, in the same way as for the simulated pseudo-random noise above. One out of the resultant 1000 data values was chosen randomly for each trial and added to the signal to emulate the realistic noise in the measured EEG data. The experiment was repeated for each of the 6203 dipole locations in the brain, and for different levels of SNR (5, 10, 15, 20, 25, 30 and infinity).

After the reconstruction was performed using sLORETA, the position of maximum standardized current density given by eq.(4) was considered as the position of the reconstructed source. The Euclidian distance between the exact source location and the reconstructed one was calculated as a measure of the localization error. As we were adding a random amount of noise, the above mentioned experiment was conducted 100 times for each dipole location $D_i$ ($\forall_i \in 1:N_V$), and the average localization error for the 100 runs was computed as the final localization error for the dipole $D_i$. The experiment was performed for both electrode setups I and II. The average localization error over the 6203 dipoles, the standard deviation of the error, the maximal and minimal localization error are listed in Table 1 with respect to each SNR level. Unsurprisingly, the localization error increases with the increase in the level of noise, however for the dipoles located in the deeper cortex areas this effect is more prominent than for the dipoles located in the superficial cortex. At the same time, Setup II (14 electrodes) was found somewhat more susceptible to noise than Setup I (19 electrodes). Figure 2 shows the mean localization error for 6203 test dipoles for different SNR levels for the two different electrode setups for simulated noise. More details about the spatial distribution of the error over the 6203 dipoles can be found in Fig. 3.

**Table 1** Average ($\overline{D}$), standard deviation ($s_D$), and standard error[6] ($s_E$) of the localization error for sLORETA in the case of a single active dipole

| SNR Level | Localization error (mm) | | | | | |
|---|---|---|---|---|---|---|
| | Setup I (Compumedics headset with 19 electrodes) | | | Setup II (Epoc headset with 14 electrodes) | | |
| | $\overline{D}$ | $s_D$ | $s_E$ | $\overline{D}$ | $s_D$ | $s_E$ |
| Infinity | 0 | 0 | 0 | 0 | 0 | 0 |
| 30 | 1.22 | 1.72 | 0.02 | 3.53 | 3.10 | 0.04 |
| 25 | 1.77 | 2.06 | 0.03 | 4.57 | 3.49 | 0.04 |
| 20 | 2.64 | 2.46 | 0.03 | 6.05 | 4.06 | 0.05 |
| 15 | 4.08 | 3.01 | 0.04 | 8.29 | 4.72 | 0.06 |
| 10 | 6.72 | 3.92 | 0.05 | 12.36 | 6.09 | 0.08 |
| 5 | 13.58 | 6.97 | 0.09 | 23.17 | 11.22 | 0.14 |

---

[6] Standard error $= \dfrac{\text{Standard deviation}}{\sqrt{\text{Number of cases}}}$

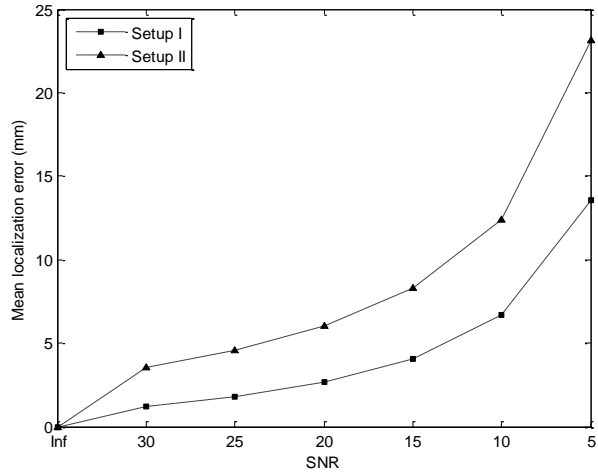

**Fig. 2** Dependence of the mean localization error on the SNR (for simulated noise).

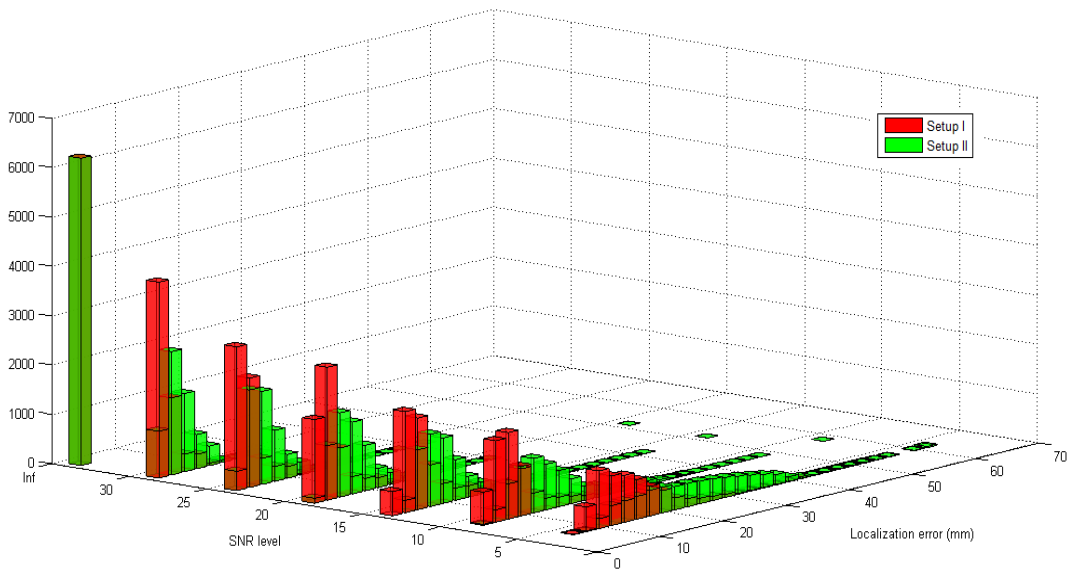

**Fig. 3** Histogram of the dipole localization errors (for simulated noise) based on 6203 test dipoles for different levels of SNR, (a) 30, (b) 25, (c) 20, (d) 15, (e) 10, (f) 5.

In the case of realistic noise we conducted experiment only for Setup II and found the results to be very similar to those in the case of the simulated noise. For SNR of 30, the mean localization error was 3.8 mm; the localization error increased with the decrease of SNR, and for SNR of 5, the mean localization error was 22.20 mm. While the mean localization errors were similar both for the simulated and the realistic noise, the standard deviation of errors was higher in the case of the realistic noise.

## 4 Width of the Reconstructed Distribution for a Single Active Source

For a particular active dipole $D_i$, and for noise-free potentials, the EEG source reconstruction was performed using sLORETA. The reconstructed signal had a broad spatial distribution covering the original source $D_i$, with the maximum strength of the reconstructed signal located at $D_i$. The reconstructed signal strength was plotted as a function of the Euclidian distance between the dipole locations $(x_i, y_i, z_i)$ and $(x_j, y_j, z_j)$ of dipoles $D_i$ and $D_j$ (for $j$=1 to 6203). While the Euclidian distances in that case can take any real value between 0 to 150 mm (size of the cortex in our model), for presentation purposes we clustered the distances within the ranges 0-4 mm, 5-9 mm and so on, averaging the dipoles strength within each cluster. Figure 4 shows an example of the plot of such a reconstruction (for Setup I) corresponding to a single active dipole source located in the centre of the neocortex. As expected, the maximum reconstructed signal strength was found at zero Euclidian distance from the simulated dipole source.

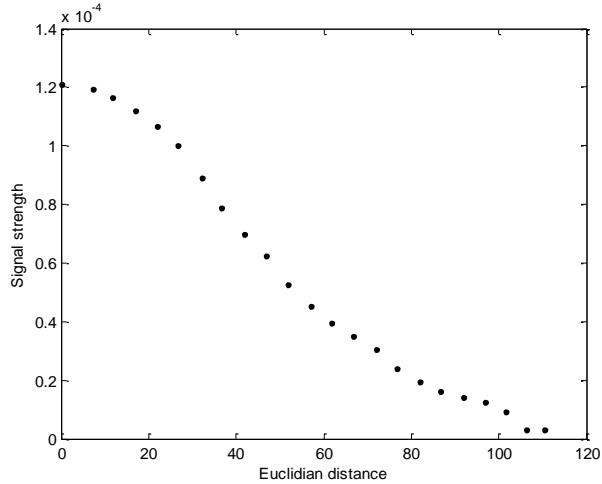

**Fig. 4** Reconstructed signal strength as a function of the Euclidian distance from the simulated source for an EEG reconstruction with a single active dipole located at the centre of the brain.

Then we computed the width of the reconstructed distribution based on its second-order integral moment, i.e. this width was defined as $2\sqrt{\sum(X_i^2 Y_i)/\sum Y_i}$, where $X_i$ and $Y_i$ represent the Euclidian distance and the signal strength, respectively. Similar calculations were performed for all dipoles $D_i$ (for $i$=1 to 6203) activated one at a time. Experiments were performed for both Setup I and Setup II.

Our findings inferred that depending on the dipole position in the neocortex, the width of the reconstructed distribution varied. Numerically, the width varied between 63 mm to 110 mm having a mean width of around 83 mm with the standard deviation of around 8 mm for Setup I. For Setup II, the width varied between 72 mm to 153 mm having a mean of around 112 mm with a standard deviation of around 14 mm. Active dipoles located in the lower occipital area led to broader width distribution of the reconstructed signal compared to dipoles located in other parts of the brain, which was presumably due to the poorer coverage of the corresponding scalp area by the electrodes. Generally, as expected, the dipoles located deeper in the brain had higher width distribution compared to the dipoles that are located closer to the electrodes (or scalp) and fall within the "good coverage area" by the electrodes. However closely located (from the scalp) dipoles could have broader distribution when they fall outside the "good coverage area" by the electrodes. Overall, these results indicate that, in agreement with previously published results, while the maximum of the reconstructed signal corresponds exactly (in the noiseless case) to the position of the single active source, the signal distribution reconstructed using sLORETA method is quite broad. Figure 5 shows the distribution of computed width over the considered 6203 dipoles, where gray levels represent the width of the reconstructed signal when that dipole was active.

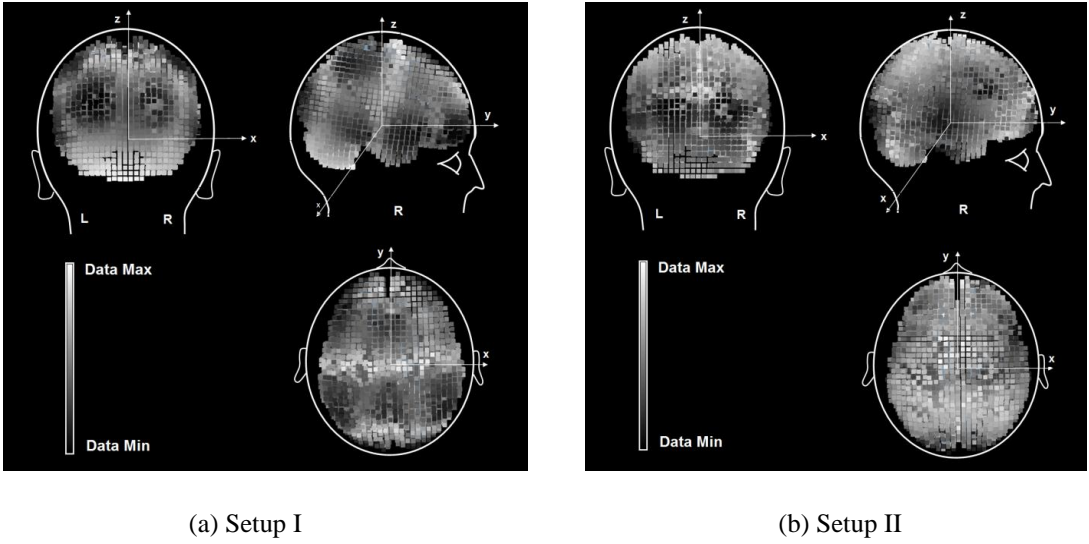

(a) Setup I  (b) Setup II

**Fig. 5** Distribution of width over 6203 dipoles, where gray levels represent the width of the reconstructed signal when that dipole was active.

## 5 Evaluating sLORETA in the Case of Simultaneously Activated Multiple Sources

In the case of two active dipoles, one dipole, which we called the primary dipole $D_{primary}$, was located in one of the three fixed locations: (a) at the centre of the brain, or (b) on the surface of the occipital lobe, or (c) on the surface of the temporal lobe. For each of the three positions of the primary dipole, we considered a secondary dipole $D_i$ ($\forall_i \in 1:N_V, D_i \neq D_{primary}$), which was activated simultaneously with the primary dipole. After the reconstruction was performed using sLORETA, positions of the local maxima of the reconstructed signal were determined.

In this case, ideally, the reconstructed signal should have two local maxima coinciding with the positions of the primary and the secondary dipoles. However because of the complex distribution of the reconstructed signal produced by sLORETA, it was observed that, depending on the radius of the area within which the local maximum search was performed, it was often possible to find one, two or several local maxima in the reconstructed signal distribution. We adopted a strategy where, while deciding whether a particular point was a local maximum or not, we considered whether it had higher values compared to all other points within a sphere with the diameter $W_i$, with $W_i$ being the width of the corresponding reconstructed signal distribution that had been calculated at the previous stage of the experimental study (with a single active dipole at the location of the centre of the sphere). Then from the detected local maxima we chose two maxima $D_{max}^1$ and $D_{max}^2$ with the highest values of the reconstructed signal (in the case of only one detected local maximum we chose $D_{max}^1 = D_{max}^2$). The location $D_{max}^1$ corresponded to the position closest to the primary source, and $D_{max}^2$ represented the other source. Then the Euclidian distances between the two reconstructed sources and the corresponding actual sources were computed, the average was taken as the localization error of sLORETA in the case of two simultaneously active sources. Experiments were conducted for both Setups I and II.

Figure 6 details the findings, where the horizontal axis represents the Euclidian distance between the primary and the secondary sources and the vertical axis represents the percentage of cases where we were able to detect 1 or 2 maxima in the reconstructed signal.

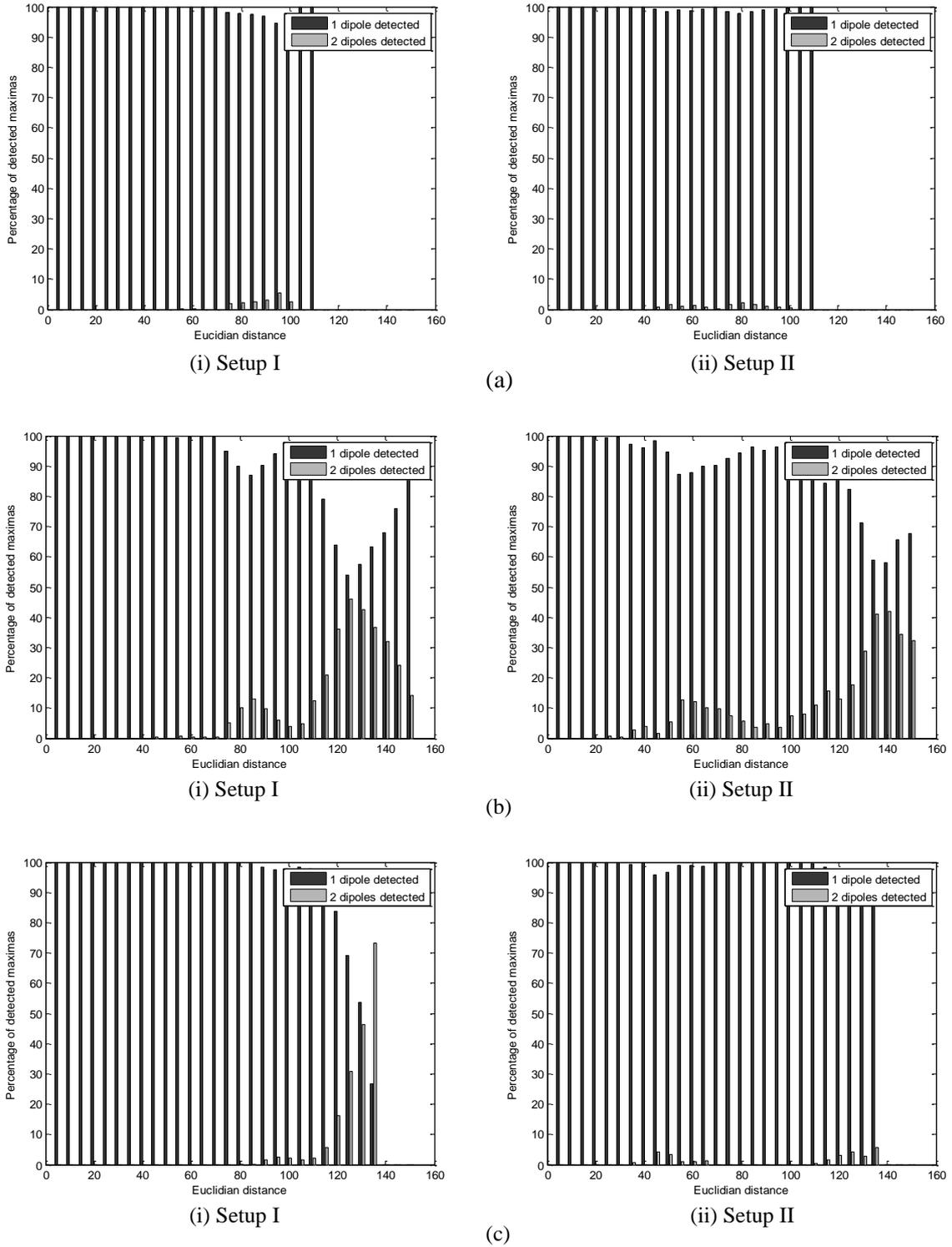

**Fig. 6** Analysis of sLORETA performance for simultaneously active sources with the primary dipole located (a) at the centre of the brain, (b) on the surface of the occipital lobe, (c) on the surface of the temporal lobe.

Table 2 summarizes the determined localization error in the case of two simultaneously active dipoles.

**Table 2** Average ($\overline{D}$), standard deviation ($s_D$), and standard error ($s_E$) of the localization error for two simultaneously active sources

| Electrode Setups | Localization error (mm) | | | | | | | | |
|---|---|---|---|---|---|---|---|---|---|
| | Primary dipole is at the centre of the brain | | | Primary dipole is at the surface of the occipital lobe | | | Primary dipole is at the surface of the temporal lobe | | |
| | $\overline{D}$ | $s_D$ | $s_E$ | $\overline{D}$ | $s_D$ | $s_E$ | $\overline{D}$ | $s_D$ | $s_E$ |
| Setup I | 36.23 | 12.10 | 0.15 | 48.93 | 23.11 | 0.29 | 46.00 | 16.58 | 0.21 |
| Setup II | 40.60 | 15.28 | 0.19 | 49.35 | 21.22 | 0.27 | 50.80 | 19.27 | 0.24 |

From these results, it is possible to conclude that when the distance between the primary and secondary dipoles was considerably larger than the width of the source distribution produced by a single active dipole, sLORETA was sometimes able to detect the two sources accurately. Note however that, in addition to the distance between the active dipoles, the respective positions of these dipoles in the brain also played a significant role for successful localization of the two maxima in the reconstructed signal.

The effect of measurement noise on the localization error produced by sLORETA in the case of simultaneously active multiple sources was also analysed. Unsurprisingly the localization error was higher in the presence of noise. For Setup I and simulated noise with SNR of 30 the localization error was about 37 mm while the primary dipole was located at the centre of the brain. For the primary dipoles on the surface of the occipital lobe and on the surface of the temporal lobe the localization errors were 49 mm and 46 mm respectively (for the same SNR). Decreased levels of SNR resulted in the increase of the localization error, e.g. for the SNR of 5, the localization errors were 40 mm, 54 mm and 49 mm for the primary dipoles positioned at the centre of the brain, on the surface of the occipital lobe, and on the surface of the temporal lobe respectively. For Setup II the computed localization errors were 41 mm, 49 mm and 50 mm respectively for the primary dipoles positioned at the centre of the brain, on the surface of the occipital lobe, and on the surface of the temporal lobe for the SNR of 30. For the SNR of 5 those values were respectively 48 mm, 52 mm and 51 mm. In the case of realistic noise with the SNR of 30 for Setup II the localization errors were 41 mm, 50 mm and 52 mm respectively for the considered dipoles positions. The SNR of 5 resulted in the localization errors of about 50 mm, 54 mm and 53 mm respectively for the dipoles positioned at the centre of the brain, on the surface of the occipital lobe, and on the surface of the temporal lobe respectively. It is worth mentioning that the obtained numerical values here were based on the average of 10 successive runs. It is obvious from these results that the localization error increased with the increase of noise level, however this increase was not dramatic since the localization error for the two simultaneously active sources without any noise was already quite large.

## 6 Discussion and Conclusions

In the present study we have quantitatively analysed and evaluated the performance of the sLORETA method for EEG source localization, considering different levels of noise in the measured electrode potentials and single and multi-dipole activation, for two different EEG electrode setups corresponding to two popular EEG headsets with 14 and 19 active measurements channels, respectively. In agreement with previous publications, the experimental results show that sLORETA produces zero localization error when there is no noise in the measured signals and only one source dipole is active at a time. Our study has confirmed that this result remains true even for the considered low-density EEG headsets. However, it was found that the localization error is non-zero if any of these two conditions is violated. We have also quantified the width of the reconstructed signal distribution for a single active source. It has been already mentioned in the literature that sLORETA produces a broad reconstructed signal; this study quantifies the width of that signal and thus specifies the lower bound for the spatial resolution of the sLORETA method for the case of the considered EEG configurations. We have performed this spatial resolution analysis first for the noise-free data, and then systematically investigated the influence of different levels of the SNR on the width of the reconstructed signal.

We have also analysed the performance of sLORETA EEG reconstruction for two simultaneously active sources. It was found in previous publications (see e.g.[17]) that when two sources were active simultaneously, they could only be separated in the reconstructed signal produced by sLORETA if they were located far apart and were of similar strength; the worst performance was also observed for parallel sources [17]. Knowing that in a real-life scenario the dipoles can be randomly oriented, the present study focused on the worst case scenario. Our computer simulation results showed that, in addition to the distance between the two active dipoles, their respective positions in the brain also affects their detectability in the reconstructed signal produced by sLORETA. Overall, the localization error in the case of two simultaneously active sources was found to be quite large (up to one third of the head diameter on average) even in the noise-free case. This localization error increased further when increasing levels of noise were added to the EEG data in our simulations, but the additional deterioration of the reconstructed results in this case was relatively modest, due to the fact that the errors were already quite large in the noise-free case.

In principle, it may be also interesting to analyse the performance of sLORETA for more than two simultaneously active sources. However, taking into account the poor performance of sLORETA for two simultaneously active sources, we can hypothesize that at least in the case of low-resolution EEG setups (with a small number of measurement channels available) sLORETA will not be able to successfully localize multiple simultaneously active sources. It may be worth, though, to study further the quantitative performance of sLORETA in the case of low-density EEG headsets and broadly distributed sources, which we plan to do in a subsequent work.

**Acknowledgment**

This project was supported by the Computational and Simulation Sciences Transformational Capability Platform of Commonwealth Scientific and Industrial Research Organisation (CSIRO), Australia, along with University of New South Wales (UNSW), Canberra, Australia. The authors would also like to thank Dr. Rajib Rana and Dr. Frank De Hoog of CSIRO Computational Informatics for their valuable comments and suggestions.

**References**

[1]     B.E. Swartz, E.S. Goldensohn (1998) Timeline of the history of EEG and associated fields. Electroencephalography and clinical Neurophysiology 106 (2): 173-176

[2]     Paul L. Nunez, Ramesh Srinivasan (2006) Electric fields of the brain: the neurophysics of EEG. Oxford University Press, USA

[3]     Chris Plummer, A. Simon Harvey, Mark Cook (2008) EEG source localization in focal epilepsy: Where are we now?. Epilepsia 49 (2): 201-218

[4]     Z. A. Acar, S. Makeig (2013) Effects of Forward Model Errors on EEG Source Localization. Brain topography 26 (3): 378-396

[5]     S. Finnigan, M. J. van Putten (2013) EEG in ischaemic stroke: quantitative EEG can uniquely inform (sub-)acute prognoses and clinical management. Clinical neurophysiology 124 (1): 10-19

[6]     T. G. Phan, T. Gureyev, Y. Nesterets, H. Ma, D. Thyagarajan (2012) Novel Application of EEG Source Localization in the Assessment of the Penumbra. Cerebrovascular diseases 33(4): 405-407

[7]     P. W. Kaplan, A. O. Rossetti (2011) EEG Patterns and Imaging Correlations in Encephalopathy: Encephalopathy Part II. Journal of Clinical Neurophysiology 28 (3): 233-251

[8]     Pascual-Marqui, Roberto Domingo (2002) Standardized low-resolution brain electromagnetic tomography (sLORETA): technical details. Methods Find Exp Clin Pharmacol 24 (Suppl D): 5-12

[9]     Roberta Grech, Tracey Cassar, Joseph Muscat, Kenneth Camilleri, Simon Fabri, Michalis Zervakis, Petros Xanthopoulos, Vangelis Sakkalis, Bart Vanrumste (2008) Review on solving the inverse problem in EEG source analysis. Journal of neuroengineering and rehabilitation 5 (1): 25

[10]    F. Natterer (1986) The mathematics of computerized tomography. Stuttgart, B G Teubner

[11]    Pascual-Marqui, Roberto Domingo (1999) Review of methods for solving the EEG inverse problem. International journal of bioelectromagnetism 1 (1): 75-86

[12]    N.G. Gençer, Samuel J. Williamson (1998) Differential characterization of neural sources with the bimodal truncated SVD pseudo-inverse for EEG and MEG measurements. IEEE Transactions on Biomedical Engineering 45 (7): 827-838

[13]    C. Silva, J. C. Maltez, E. Trindade, A. Arriaga, E. Ducla-Soares (2004) Evaluation of L1 and L2 minimum norm performances on EEG localizations. Clinical neurophysiology 115 (7): 1657-1668

[14]    Leo K. Cheng, John M. Bodley, Andrew J. Pullan (2003) Comparison of potential-and activation-based formulations for the inverse problem of electrocardiology. IEEE Transactions on Biomedical Engineering 50 (1): 11-22

[15]    Hesheng Liu, Paul H. Schimpf, Guoya Dong, Xiaorong Gao, Fusheng Yang, Shangkai Gao (2005) Standardized shrinking LORETA-FOCUSS (SSLOFO): a new algorithm for spatio-temporal EEG source reconstruction. IEEE Transactions on Biomedical Engineering 52(10): 1681-1691

[16]    Hesheng Liu, Xiaorong Gao, Paul H. Schimpf, Fusheng Yang, Shangkai Gao (2004) A recursive algorithm for the three-dimensional imaging of brain electric activity: shrinking LORETA-FOCUSS. IEEE Transactions on Biomedical Engineering 51 (10): 1794-1802

[17]    Michael Wagner, Manfred Fuchs, Jörn Kastner (2004) Evaluation of sLORETA in the presence of noise and multiple sources. Brain Topography 16 (4): 277-280

[18]    Lei Ding, Bin He (2008) Sparse source imaging in electroencephalography with accurate field modelling. Human brain mapping 29 (9): 1053-1067

[19]    M. S. Hamalainen, J. Sarvas (1989) Realistic conductivity geometry model of the human head for interpretation of neuromagnetic data. IEEE Transactions on Biomedical Engineering 36 (2): 165-171

[20]    EPOC Features,WWW Document, (http://www.emotiv.com/epoc/)

[21] Compumedics,WWW Document, (http://www.compumedics.com/products.asp?p=39)

[22] M.S. Hämäläinen, R.J. Ilmoniemi (1984), Interpreting measured magnetic fields of the brain: estimates of current distributions Tech. Rep. TKK-F-A559, Helsinki University of Technology, Espoo

[23] Dale AM, Liu AK, Fischl BR, Buckner RL, Belliveau JW, Lewine JD, Halgren E. (2000) Dynamic statistical parametric mapping: combining fMRI and MEG for high resolution imaging of cortical activity. Neuron 26: 55-67

[24] G. Van Hoey, Bart Vanrumste, M. D'Havé, Rik Van de Walle, Ignace Lemahieu, Paul Boon (2000). Influence of measurement noise and electrode mislocalisation on EEG dipole-source localisation. Medical and Biological Engineering and Computing 38 (3): 287-296

[25] G. Van Hoey, Jeremy De Clercq, Bart Vanrumste, Rik Van de Walle, Ignace Lemahieu, Michel D'Havé, Paul Boon (2000) EEG dipole source localization using artificial neural networks. Physics in medicine and biology 45 (4): 997-1011